\documentclass[onecolumn,amsfonts,amssymb,amsmath,floatfix]{article}
\usepackage[dvips]{graphicx}
\usepackage{natbib}
\usepackage{bm}

\title{Scalar potential model of spiral galaxy H\,{\Large\bf I} rotation curves and rotation curve asymmetry}

\author{J.C. Hodge$^{1}$\thanks{E-mail:jch9496@blueridge.edu}\thanks{Visiting from XZD Corp., 16 Hosta Ln., Brevard, NC, 28712, E-mail:{scjh@citcom.net}}\\
$^{1}${Blue Ridge Community College, 100 College Dr., Flat Rock, NC, 28731-1690}}

\date{\today}

\begin{document}

\maketitle

\begin{abstract}
A scalar potential model (SPM) was developed from considerations of galaxy clusters and of redshift.  The SPM is applied to H\,{\sc i} rotation curves (RCs) and RC asymmetry of spiral galaxies.  The resulting model adds the force of a scalar potential of the host galaxy and of neighboring galaxies to the Newtonian rotation velocity equation.  The RC is partitioned radially into regions.  The form of the equation for each parameter of each region is the same with differing proportionality constants.  Integer values of each equation are determined empirically for each galaxy.  Among the sample galaxies, the global properties of galaxies of B band luminosity, of position, and of orientation determine the RC and RC asymmetry.  The Source of the scalar field acts as a monopole at distances of a few kpc from the center of spiral galaxies.  The scalar potential field causes Newtonian mechanics to considerably underestimate the mass in galaxies, which is the ``missing mass problem''.  The SPM is consistent with RC and RC asymmetry observations of the sample spiral galaxies.
\end{abstract}

Galaxies: kinematics and dynamics -- Galaxies: fundamental parameters -- Galaxies: interactions -- Galaxies: spiral

\section[SPM of RCs and RC asymmetry]{Introduction}

The discrepancy between the Newtonian estimated mass in spiral galaxies and the observation of star and gas kinematics is well established.  The rotation curve (RC) relates the square of the tangential rotation velocity $v $ (km s$^{-1}$) of particles in orbit around spiral galaxies to their galactocentric radius $R$ (kpc).  The RC is measured along the major axis.  Traditionally, the focus has been on accounting for H\,{\sc i} RCs that are flat in the outer region immediately beyond the knee (OR).  However, observations also include rising RCs, declining RCs, an abrupt change in slope at the extreme outer region (EOR) in many galaxies, and rotational asymmetry with non-relativistic velocities.  \citet{batt,fer5,ghez,sofu,sofu2}; and \citet{taka} provide a summary of the knowledge of RCs.  The RC is an excellent tool to evaluate galaxy models.  

\begin{figure}
\centering 
\includegraphics[width=0.5\textwidth]{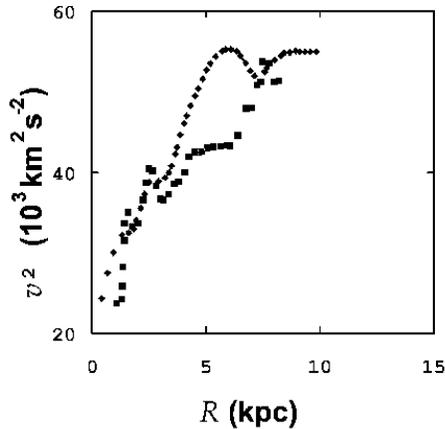}
\caption{Plots of the square of the rotation velocity $v^2$ (10$^3$ km$^2 \,$s$^{-2}$) versus galactocentric radius $R$ (kpc) of the H\,{\sc i} RC (filled diamonds) and H$\alpha$ RC (filled squares) for NGC4321 \citep{semp}. }
\label{fig:N4321}
\end{figure}

The RC differs for different particles.  For example, the H\,{\sc i} RC and the RCs of stars as shown by the H$\alpha$ line for {NGC~4321} \citep{semp} differ in the rapidly rising region before the knee (RR) and approach each other in the OR as shown in Fig.~\ref{fig:N4321}. 

The stars near the center of galaxies have rotation velocities more than an order of magnitude larger than $v$ at the knee of the H\,{\sc i} RC.  The decline of the stellar RCs from the center is Keplerian \citep{ghez,fer5}.  The $v$ continues to decline to the radius $R_\Delta$ (kpc) at which the rotation velocity $v_\Delta$ is at a minimum.  The $v_\Delta$ is approximately an order of magnitude smaller than $v$ at the knee of the H\,{\sc i} RC. 

The H\,{\sc i} RC rarely extends inward to near $R_\Delta$.  When drawing the H\,{\sc i} RC, a straight line is often drawn from the innermost measured point through the center of the galaxy.  This practice is inconsistent with stellar observations~\citep{scho} and is incorrect~\citep{sofu2}.

The evidence that the cluster environment affects RC shape appears contradictory \citep{whit,dale}.  These studies concentrated on H$\alpha$ RCs.  The H\,{\sc i} RC usually extends farther outward than H$\alpha$ RCs.  Therefore, the H\,{\sc i} RCs are more sensitive to neighboring galaxy's effects. 

As larger samples of RCs have been measured, researchers attempted to produce a quantitative description of an empirically derived, average RC.  \citet{pers} claimed the existence of a universal RC (URC) for a large sample revealed by the feature that only luminosity of the host galaxy dictated the RC.  Independent work has confirmed much of the URC \citep[and references therein]{cati}.  Other researchers have discussed the inadequacy of the average RC (e.g. \citealt{cour,garr,willick}).  

\citet{rosc} used a dynamical partitioning process and found that the dynamics in the outer part of optical RCs are constrained to occupy one of four \emph{discrete dynamical classes}.  The classes are determined by the absolute magnitude, surface brightness, and a parameter for each optical RC that is an exponent of the radius at which $v$ is measured. 

Galaxies with a small asymmetry in the RC of a few km~s$^{-1}$ are well known \citep{shan}.  Large RC asymmetries are less appreciated \citep{jog}.  Because the observational data from each side of a galaxy RC are generally averaged, only highly asymmetric cases are recognized.  RC asymmetry appears to be the norm rather than the exception \citep{jog}.  \citet{wein} and \citet{jog97} proposed the implied mass asymmetry is due to an imposed lopsided potential caused by galaxy interaction.  \citet{dale} found RC asymmetry of early type galaxies falls by a factor of two between the inner and outer regions of clusters.  The formation, evolution, and long term maintenance of galactic, kinematical asymmetry remains a mystery. 

With the improved H\,{\sc i} measuring equipment in the 1990's, the H\,{\sc i} RC observations detected an abrupt change in the EOR in many galaxies (see figures in \citealt{cati}).  This characteristic is ignored in current models.  For example, \citet{pizz05} considered this factor as a reason to exclude galaxies from the sample.

Spectra coming from HII regions are correlated systematically with $R$ and little else \citep[pp. 516-522]{binn}.  The interstellar abundances of metals in a disk galaxy decrease with increasing radius \citep{tadr}.  Also, the absolute B band magnitude $M_\mathrm{B}$ (mag.) of a galaxy is correlated with the metallicity obtained by extrapolating [O/H] within the disk to the galactic center.  Low luminosity galaxies tend to be metal poorer than high luminosity galaxies. 

If most of the matter of a galaxy is in the bulge region, classical Newtonian mechanics predicts a Keplerian declining RC in the disk.  The observation of rising RCs in the RR and of rising, flat, or declining RC's in the OR poses a perplexing problem.  Postulates to model the RCs have included the existence of a large mass $M_\mathrm{DM}$ of non-baryonic, dark matter (DM) and dark energy; the variation of the gravitational acceleration $R$; the variation of the gravitational constant $G$ with $R$; and the existence of an added term to the Newtonian gravitation equation opposing gravitational attraction. 

Repulsive DM has been proposed to provide a repulsive force that has the form of an ideal, relativistic gas \citep{good}.  In DM models the rising RCs of low surface brightness galaxies (LSBs) require larger relative amounts of DM than flat or declining RCs.  This contradiction to the successful Tully-Fisher relation (TF) \citep{tull77} is well established \citep{sand}.  Another mystery of the DM paradigm is the apparent strong luminous-to-dark matter coupling that is inferred from the study of RCs \citep{pers}.

Currently, the most popular modified gravity model is the Modified Newtonian Dynamics (MOND) model~\citep[and references therein]{bott,sand}.  MOND suggests gravitational acceleration changes with galactic distance scales by a fixed parameter related to the mass-to-luminosity ratio with units of acceleration.  MOND appears limited to the disk region of spiral galaxies.  MOND requires the distance for {NGC 2841} and {NGC 3198} to be considerably larger than the Cepheid-based distance $D_\mathrm{a}$ (Mpc)~\citep{bott}.  Appling MOND to cosmological scales is being investigated~\citep{beke,sand2}.  MOND may represent a \emph{effective} force law arising from a broader force law~\citep{mcga}.  

The possible variation of $G$ or $\vert \dot{G}/G \vert \sim H_\mathrm{o}$, where $ H_\mathrm{o}$ is the Hubble constant, were first proposed by Dirac.  This was followed by the Brans-Dicke theory of gravity that postulated $G$ behaves as the reciprocal of a scalar field whose source is all matter in the universe~\citep{narl}.  More recently \citet[and references therein]{brown} have proposed a variation of general relativity in which $G$ varies with $R$.  This model has been applied to RCs \citep{brown,moff} and the postulated gravitational radiation of binary pulsars.  This model has results similar to the MOND model.  However, it uses a distance $D$ (Mpc) for {NGC 2841} of 9.5 Mpc rather than $D_\mathrm{a} =$14.07~Mpc \citep{macr}.

Alternate models of the RC posit the existence of another term in the Newtonian equation resulting from a scalar field of a ``fifth force'', a scalar field coupled to matter, or a static\footnote{Static in analogy to an electrostatic field caused by charged particles.} scalar field coupled to DM.  \citet[and references therein]{padm} explored the possibility of a scalar field acting as dark matter.  The usual ``fifth force'' models posit a Yukawa like potential~\citep{bert}.  \citet{fay} postulated minimally coupled and massive scalar fields are responsible for flat RCs, only.  \citet{mbel} suggested a real scalar field, minimally coupled to gravity sourced by baryonic and dark matter may also reproduce flat and rising RCs with the assumption that the dark halo mass density dominates the baryonic mass density.  The free parameters are a turnover radius $r_\mathrm{o}$, the maximum rotation velocity, which is at $r_\mathrm{o}$, and an {\emph{integer}}.  If a natural theoretical explanation could be found for the relation between $r_\mathrm{o}$ and the positive integer, the integer may be the only free parameter.  The postulate of an unknown ``fifth force'' is as likely as the postulate of unknown and undetected DM. 

The scalar potential model (SPM) was created to be consistent with the observation of the morphology-radius relation of galaxies in clusters, of the intragalactic medium of a cluster of galaxies, and of the flow of matter from spiral galaxies to elliptical galaxies~\citep{hodg}.  The SPM was applied to redshift observations and found a qualitative explanation of discrete redshift observations.  The SPM suggests the existence of a massless scalar potential $\rho \propto R^{-1}$ derived from a heat differential equation.  Physically, the heat equation requires a flow\footnote{A flow of energy is in contrast to a static field.  An analogy is the flow of gas or fluid from Sources to Sinks.} of energy from Sources to Sinks to form the potential field.  That the $\vec{\nabla}\rho$ field acts like a ``wind'' was suggested.  Several differences among galaxy types suggest that Sources of $\rho$ are located in spiral galaxies and that Sinks of $\rho$ are located in elliptical, lenticular, and irregular galaxies.  The Source forming a galaxy leads to the proportionally of the Source strength and emitted radiation (luminosity).  Therefore, the total mass of a galaxy is related to the luminosity of a galaxy.  A cell structure of galaxy groups and clusters was proposed with Sinks at the center and Sources in the outer shell of the cells.  The cell model is supported by the data and analysis of \citet{aaro2,cecc,huds,lilj}; and \citet{rejk}.  Because the distance between galaxies is larger than the diameter of a galaxy, the Sources were considered as point (monopole) Sources.  

The force $\vec{F}_\mathrm{s}$ (dyne) of the $\rho$ field that acts on matter is 
\begin{equation}
\vec{F}_\mathrm{s} = G_\mathrm{s} m_\mathrm{s} \vec{\nabla} \rho
\label{eq:Fs} \,,
\end{equation}
where (1) the $G_\mathrm{s}$ is a proportionality constant; (2) $m_\mathrm{s}$ is the property of particles on which $\vec{F}_\mathrm{s}$ acts; and (3) $\rho$ is the sum of the effects of all galaxies,
\begin{equation}
\rho  =  K_\epsilon \sum_{i=1}^{N_\mathrm{source}} \frac{\epsilon_i}{r_{i}} + K_\eta  \sum_{l=1}^{N_\mathrm{sink}} \frac{\eta_l}{r_{l}}
\label{eq:13b} \,,
\end{equation}
where $\epsilon_i$ and $\eta_l$ are numbers representing the strength of the $i^{th}$ Source and $l^{th}$ Sink, respectively, $ K_\epsilon $ and $ K_\eta $ are proportionality constants, and $r_{i} $ (Mpc) and $r_{l}$ (Mpc) are the distance from a Source and a Sink, respectively, to the point at which $\rho $ is calculated, $\epsilon>0$, $\eta<0$, and $N_\mathrm{source}$ and $N_\mathrm{sink}$ are the number of Sources and Sinks, respectively, used in the calculation. 

\citet{hodg} suggested the $m_\mathrm{s}$ property of matter is the cross section of the particle; the $\epsilon \propto M_{\mathrm{t} \epsilon}$, where $ M_{\mathrm{t} \epsilon}$ is the total mass of the Source galaxy; and the $\eta \propto M_{\mathrm{t} \eta}$, where $ M_{\mathrm{t} \eta}$ is the total mass of the Sink galaxy. 

This Paper further constrains the SPM to describe another set of observations inconsistent with present models.  The SPM finds the RC in the OR may be rising, flat, or declining depending on the galaxy and its environment.  The Source of the scalar field acts as a monopole at distances of a few kpc from the center of spiral galaxies.  In addition to fitting the RR and OR parameters, the SPM is consistent with EOR and asymmetry observations.  In section~2, the model is discussed and the SPM $v^2$ calculation equation is developed.  The resulting model is used to calculate RC parameters in Section~3.  The discussion and conclusion are in Section~4.  

\section[SPM of RCs and RC asymmetry]{Spiral galaxy model}

The coordinate system center was placed at the sample galaxy's kinematical center and was aligned to our line of sight.  The SPM posits the simplified Newtonian calculation of $v$ of a test particle is the first approximation of $v$ in an isolated galaxy, that observed asymmetries and deviations from the simplified Newtonian calculation in spiral galaxies are the result of the influence of neighbor galaxies, and that the $v^2$ of a particle in orbit of a spiral galaxy is the sum of the effects of $F_\mathrm{s}$ and the gravitational force $F_\mathrm{g}$.  Therefore,
\begin{equation}
v^2 = \frac{G M}{R} -  G_\mathrm{s} \frac{m_\mathrm{s}}{m_\mathrm{\iota}}  \frac{L}{R} - \vec{R}  \bullet \vec{a}_\mathrm{o} + \vert \vec{R}  \times \vec{a}_\mathrm{o} \vert 
\label{eq:51} \,,
\end{equation}
where (1) the Newtonian simplifications of circular motion ($\ddot{R} =0$) \citep[p. 725]{binn} and a spherical mass density field are assumed, (2) the mass of the test particle is assumed to be constant over time, (3) $M$ (M$_\odot$) is the mass inside the sphere of radius $R$ from Newton's spherical property, (4) the inertial mass $m_\mathrm{\iota}$ equals gravitational mass $m_\mathrm{g}$ of the test particle \citep{will}, (5) the $L$ term is due to the $F_\mathrm{s}$ of the host galaxy, which also derives from a spherically symmetric $\rho$ field; (6) $L=K_\mathrm{\epsilon} \epsilon = 10^{-0.4 M_\mathrm{B}}$ erg s$^{-1}$ for Source galaxies or $L = K_\mathrm{\eta} \eta = -2.7 \,\times \, 10^{-0.4 M_\mathrm{B}}$ erg s$^{-1}$ for Sink galaxies \citep{hodg}; (7) $\vert \, \vert$ indicates absolute value; and (8) $\vec{a}_\mathrm{o}$ (km~s$^{-2}$) is the acceleration caused by neighboring galaxies, 
\begin{equation}
\vec{a}_{\mathrm{o}} = \frac{G_\mathrm{s} m_\mathrm{s}}{m_\mathrm{\iota}} \vec{\nabla}\rho
\label{eq:51ao} \,,
\end{equation}
where the number of galaxies exclude the host galaxy.  Note that no assumption about the significance of $\vec{a}_{\mathrm{o}}$ has been made.  

Because $v$ is measured only along the major axis in the region under consideration~\cite[p. 725]{binn} and if the $\vec{\nabla} \rho$ field is approximately uniform across a galaxy, Eq.~(\ref{eq:51}) becomes 
\begin{equation}
v^2 = G \frac{M}{R_\mathrm{major}} -  \frac{ G_\mathrm{s} m_\mathrm{s}}{m_\mathrm{\iota}} \frac{ L }{R_\mathrm{major}} + \vert \vec{K} \bullet \vec{a}_\mathrm{o} \vert R_\mathrm{major} 
\label{eq:51c} \,,
\end{equation}
where $\vec{K}$ (km kpc$^{-1}$) is a constant vector and $R_\mathrm{major}$ (kpc) is the galactocentric radius along the major axis.

In Fig.~\ref{fig:N4321} the H\,{\sc i} RC at lower radius $R_\mathrm{rr}$ (kpc) in the RR has two scalloped shapes that suggests spherically symmetric shells of matter.  Also, the H$\alpha$ RC rapidly increases, peaks, and then declines at the beginning of each shell.  The H$\alpha$ lines are generally formed in excited interstellar gas.  In the disk region of a galaxy, the gas is usually excited by hot stars \citep{binn}.  Because the $m_\mathrm{s}/m_\mathrm{\iota}$ factor must be different for different matter types, each shell has a different metallicity star type.    Because the H$\alpha$ RC approaches the H\,{\sc i} RC in the disk region such as plotted in Fig.~\ref{fig:N4321} with hot, hydrogen burning stars, the $ m_\mathrm{s}/m_\mathrm{\iota}$ factor must be the same for H\,{\sc i} and hydrogen stars.  This suggests the $ m_\mathrm{s}/m_\mathrm{\iota}$ factor varies by element type and acts on atoms at the largest.  The metallicity - radius relation follows.

The $ m_\mathrm{s}/m_\mathrm{\iota}$ ratio of stars is changing through changing elemental composition by nucleosynthesis in addition to accretion and emission of matter.  Therefore, the H\,{\sc i} RC is preferred to trace the forces influencing a galaxy outside the bulge.  Because only the H\,{\sc i} RC is considered in the calculations herein, the units used were \mbox{$G_\mathrm{s} m_\mathrm{s}/m_\mathrm{\iota} =1$ kpc~km$^2$~s$^{-1}$~erg$^{-1}$}.

\section[SPM of RCs and RC asymmetry]{Results}

\subsection{Sample}

The elliptical, lenticular, irregular, and spiral galaxies used in the calculations were the same as used in \citet{hodg}.  That is, they were selected from the NED database\footnote{The Ned database is available at http://nedwww.ipac.caltech.edu.  The data were obtained from NED on 5 May 2004.}.  The selection criteria were that the heliocentric redshift $z_\mathrm{mh}$ be less than 0.03 and that the object be classified as a galaxy.  The parameters obtained from the NED database included the name, equatorial longitude $E_\mathrm{lon}$ (degrees) for J2000.0, equatorial latitude $E_\mathrm{lat}$ (degrees) for J2000.0, morphology, the B-band apparent magnitude $m_\mathrm{b}$ (mag.), and the extinction $E_{\mathrm{xt}}$ (mag.) as defined by NED.  The galactocentric redshift $z$ was calculated from $z_\mathrm{mh}$. 

The 21-cm line width $W_\mathrm{20}$ (km~s$^{-1}$) at 20 percent of the peak, the inclination $i_\mathrm{n}$ (arcdegrees), the morphological type code ``t'', the luminosity class code ``lc'', and the position angle $P_\mathrm{a}$ (arcdegrees) for galaxies were obtained from the LEDA database\footnote{The LEDA database is available at http://leda.univ-lyon.fr.  The data were obtained from LEDA on 5 May 2004.} when such data existed. 

\begin{figure*}
\centering 
\includegraphics[width=\textwidth]{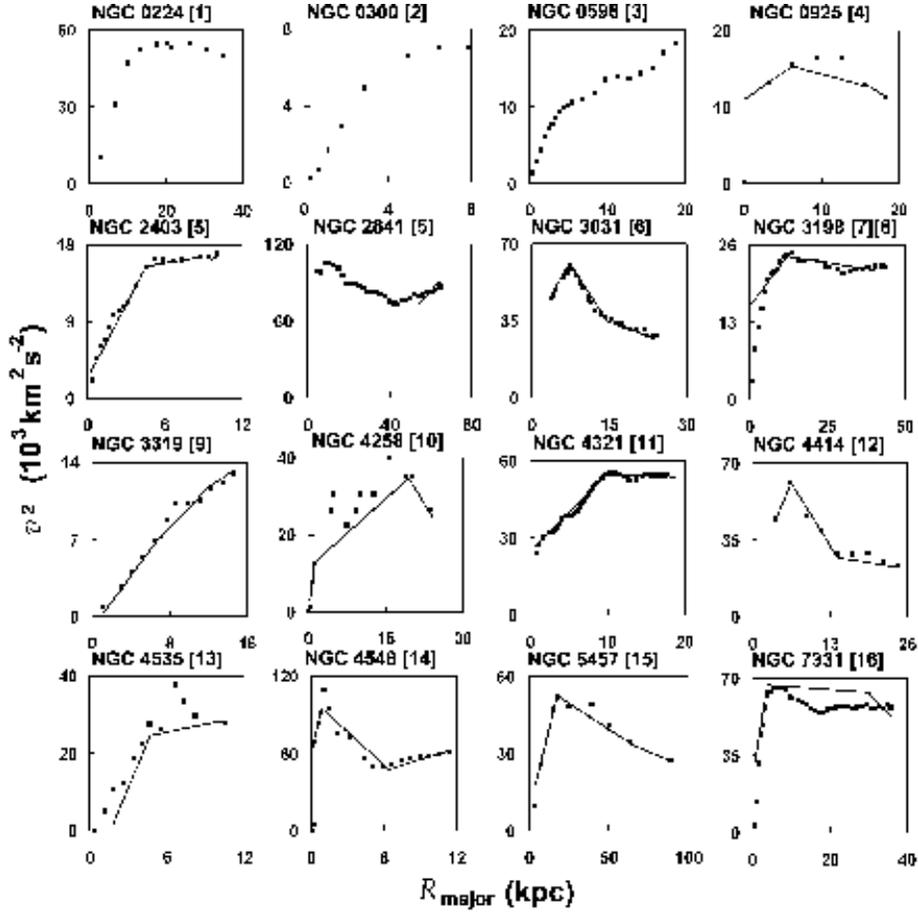}
\caption{Plots of the square of the H\,{\sc i} rotation velocity $v^2$ (10$^3$ km$^2 \,$s$^{-2}$) versus galactocentric radius $R_\mathrm{major}$ (kpc) along the major axis.  The straight lines mark the application of the derived equations to the RCs of the select galaxies.  The application of the derived equations to {NGC 0224}, {NGC 0300}, and {NGC 0598} were omitted because these galaxies lacked a $\vert \vec{K} \bullet \vec{a}_\mathrm{o} \vert$ value.  The references for the RCs are noted in brackets and are as follows: [1]\citet{gott}, [2]\citet{car3}, [3]\citet{corb}, [4]\citet{krum}, [5]\citet{bege}, [6]\citet{rots}, [7]\citet{vana}, [8]\citet{bosm3}, [9]\citet{moor}, [10]\citet{vana2}, [11]\citet{semp}, [12]\citet{brai}, [13]\citet{chin}, [14]\citet{voll}, [15]\citet{huch}, and [16]\citet{mann}. }
\label{fig:2}
\end{figure*}

The selection criteria for the ``select galaxies'' were that a H\,{\sc i} RC was available in the literature and that $D_\mathrm{a}$ was available from \citet{free} or \citet{macr}.  Data for the 16 select galaxies are shown in Table \ref{tab:galdata} and their H\,{\sc i} RCs are plotted in Fig.~\ref{fig:2}.  The $L$ for the select galaxies was calculated using $D_\mathrm{a}$, $m_\mathrm{b}$, and $E_{\mathrm{xt}}$.  The selection criteria for the other sample galaxies were that a H\,{\sc i} RC was available in the literature and that the $L$ could be calculated using the TF method with the constants developed in \citet{hodg}. 

\begin{table}
\begin{minipage}[]{\columnwidth}
\scriptsize
\caption{Data for the select galaxies.}
\label{tab:galdata}
\centering
\begin{tabular}{llllcl}
\hline
Galaxy &morphology\footnote{Galaxy morphological from the NED database.} &t\footnote{Galaxy morphological type code from the LEDA database.}&lc\footnote{Galaxy luminosity class code from the LEDA database.}&RC\footnote{Galaxy's H{\tiny{I}} RC type according to slope in the OR.  R is rising, F is flat, and D is declining.}&$D_\mathrm{a}$\footnote{The distance $D_\mathrm{a}$ (Mpc) to the galaxy from~\cite{free} unless otherwise noted.} \\
\hline
{NGC 0224} &SA(s)b  &3 &2 &F &\phantom{0}0.79\\
{NGC 0300} &SA(s)d &6.9 &6 &R &\phantom{0}2.00\\
{NGC 0598} &SA(s)cd &6 &4 &R &\phantom{0}0.84\\
{NGC 0925} &SAB(s)d HII&7 &4 &D &\phantom{0}9.16\\
{NGC 2403} &SAB(s)cd &6 &5 &F &\phantom{0}3.22\\
{NGC 2841} &SA(r)b;LINER Sy &3 &1 &F &14.07\footnote{The $D_\mathrm{a}$ is from~\cite{macr}.}\\
{NGC 3031} &SA(s)ab;LINER Sy1.8 &2.4 &2 &D &\phantom{0}3.63\\
{NGC 3198} &SB(rs)c &5.2 &3 &F &13.80\\
{NGC 3319} &SB(rs)cd;HII &5.9 &3.3 &R &13.30\\
{NGC 4258} &SAB(s)bc;LINER Sy1.9&4&3.9&D&\phantom{0}7.98\\
{NGC 4321} &SAB(s)bc;LINER HII&4 &1 &D &15.21\\
{NGC 4414} &SA(rs)c? LINER&5.1&3.6&D&17.70\\
{NGC 4535} &SAB(s)c &5 &1.9 &D &15.78\\
{NGC 4548}&SBb(rs);LINER  Sy&3.1&2&D&16.22\\
{NGC 5457}&SAB(rs)cd&5.9&1&D&\phantom{0}6.70\\
{NGC 7331} &SA(s)b;LINER &3.9 &2 &F &14.72\\
\hline
\end{tabular}
\end{minipage}
\end{table}

This select galaxy sample has LSB, medium surface brightness (MSB), and high surface brightness (HSB) galaxies; has LINER, Sy, HII, and less active galaxies; has galaxies that have excellent and poor agreement between the distance $D_{tf}$ (Mpc) calculated using the TF relation and $D_\mathrm{a}$; has a $D_\mathrm{a}$ range of from 0.79 Mpc to 17.70 Mpc; has field and cluster galaxies; and has galaxies with rising, flat, and declining RCs. 

\subsection{First approximation}

When modeling an RC, the knee or turnover of an H\,{\sc i} RC is often drawn smooth and relatively rounded.  As depicted in Fig.~\ref{fig:2} most of RCs of the selected galaxies have an abrupt change of slope at the turnover.  

The end of the RR was found by fitting a least squares straight line to the first three data points, if available, inward and immediately before the knee of the RC with a 0.99 or higher correlation coefficient.  {NGC 2841} lacked data points to identify the RR.  {NGC 4414} had only two data points before the knee and a significant decline in $v^2$ to indicate the RR.  The linear relation suggests a cylindrical distribution of matter in the RR near the knee.

At $R_\Delta$, $v_\Delta \approx 0$ that implies the effective mass $ \approx 0$ inside a sphere with a $R_\Delta$ radius.  That is, from Eq.~(\ref{eq:51c}) \mbox{$ M_\Delta \approx G_\mathrm{s} m_\mathrm{s} L / G m_\mathrm{\iota}$, where $M_\mathrm{\Delta}$} is the mass within $R_\Delta$ at which $\mathrm{d}v^2/\mathrm{d}R =0$ and at which the RR begins. 

The mass in an elemental volume d$R$ in the RR near the turnover was modeled as a cylinder shell of small height $H_\mathrm{rr}$~\cite[p. 724]{binn} and with density $D_\mathrm{rr}$ at $R_\mathrm{rr}$.  The mass $M_\mathrm{rr}$ within a sphere with a $R_\mathrm{rr}$ radius is
\begin{equation}
M_\mathrm{rr}= \left( D_\mathrm{rr} H_\mathrm{rr} 2 \pi \right) R_\mathrm{rr}^2 + M_\mathrm{\Delta} + M_\mathrm{c}
\label{eq:62} \,,
\end{equation}
where $ M_\mathrm{c}$ is a small correction term, the mass in the thin cylindrical RR shell $\approx$ the mass in the spherical shell with the same radial parameters, and the $M_\mathrm{\Delta}$ is distributed in a much thicker volume, which may be spherical.  The components of $ M_\mathrm{c}$ are $- \left( D_\mathrm{rr} H_\mathrm{rr} 2 \pi \right) R_\Delta^2$, the small amount of matter indicated by $v_\Delta \neq 0$, and the variation of the mass in the RR from a cylindrical distribution.  

Inserting Eq.~(\ref{eq:62}) into Eq.~(\ref{eq:51c}) for the RR yields 
\begin{eqnarray}
v_\mathrm{rr}^2 &=& \left(\frac{m_\mathrm{g}}{m_\mathrm{\iota}} G D_\mathrm{rr} H_\mathrm{rr} 2 \pi \right) R_\mathrm{rr} \nonumber \\*
& & + \frac{G (M_\mathrm{\Delta} +M_\mathrm{c}) - m_\mathrm{s} G_\mathrm{s} L / m_\mathrm{\iota} }{ R_\mathrm{rr}} \nonumber \\*
& & - \vert \vec{K} \bullet \vec{a}_\mathrm{o} \vert R_\mathrm{rr}  
\label{eq:64} \,,
\end{eqnarray}
where $v_\mathrm{rr}$ (km~s$^{-1}$) is the rotation velocity of H\,{\sc i} in the RR at $R_\mathrm{rr}$.

By averaging from side-to-side of the galaxy, the effect of the $\vert \vec{K} \bullet \vec{a}_\mathrm{o} \vert$ term was reduced and was considered insignificant for the first approximation compared to the $D_\mathrm{rr}$ term near the knee.  The $\vec{F}_\mathrm{g}$ and the $\vec{F}_\mathrm{s}$ offset each other at $R_\Delta$ to produce a minimum $v$ that then rises in the RR. Because $v_\Delta \approx 0$ and the RCs near the knee are well fit by the straight lines, the \mbox{$[G (M_\mathrm{\Delta r} + M_\mathrm{c}) - m_\mathrm{s} G_\mathrm{s} L m_\mathrm{\iota}^{-1} ]/ R_\mathrm{rr}$} term is small and nearly constant in the RR compared to the $D_\mathrm{rr}$ term near the knee of the RC.  Equation~(\ref{eq:64}) suggest $ v_\mathrm{rr}^2 $ depends on galaxy parameters $D_\mathrm{rr}$ and $H_\mathrm{rr}$.  Therefore, the first approximation is 
\begin{equation}
\frac{v_\mathrm{rr}^2}{10^3 \, \mathrm{km}^2 \, \mathrm{s}^{-2}}= S_\mathrm{rr} \frac{R_\mathrm{rr}}{\mathrm{kpc}} + I_\mathrm{rr} 
\label{eq:65} \,,
\end{equation}
where $S_\mathrm{rr}$ and $I_\mathrm{rr}$ are the slope and intercept of the linear $v_\mathrm{rr}^2 - R_\mathrm{rr}$ relationship, respectively.  

As $R_\mathrm{rr}$ increases, $m_\mathrm{s} /m_\mathrm{\iota}$ of the predominant mass increases and the predominant particles have decreasing density.  Decreasing the amount of mass in a given cylindrical shell causes the end of the RR and the beginning of the transition region (TR) between the RR and OR.  Because $\epsilon \propto M_\mathrm{t \epsilon}$ and $\epsilon \propto M_\Delta$, the mass $M_\mathrm{rrmax}$ within the maximum RR radius $R_\mathrm{rrmax}$ (kps) $\propto \epsilon \propto L$ of Eq.~(\ref{eq:51c}).  The rotation velocity $v_\mathrm{rrmax}$ (km s$^{-1}$) of the galaxy at $R_\mathrm{rrmax}$ can be used to compare parameters among galaxies.  Because this change is caused by the change in elemental types that depends on the $\epsilon$ and, hence, $L$ of a galaxy, the end of the RR is occurring under similar conditions in all galaxies.  Thus, the $v^2_\mathrm{rrmax}$ and $R_\mathrm{rrmax}$ are comparable parameters among galaxies.  Posit, as a first approximation, 
\begin{equation}
\frac{v^2_\mathrm{rrmax}}{10^3 \mathrm{km}^2 \, \mathrm{s}^{-2}} = S_\mathrm{a} \frac{L}{10^8\, \mathrm{erg\,s}^{-1}} + I_\mathrm{a}
\label{eq:57m} \,,
\end{equation}
where $S_\mathrm{a}$ and $ I_\mathrm{a}$ are the slope and intercept of the linear relation, respectively.

\begin{figure}
\centering 
\includegraphics[width=0.5\textwidth]{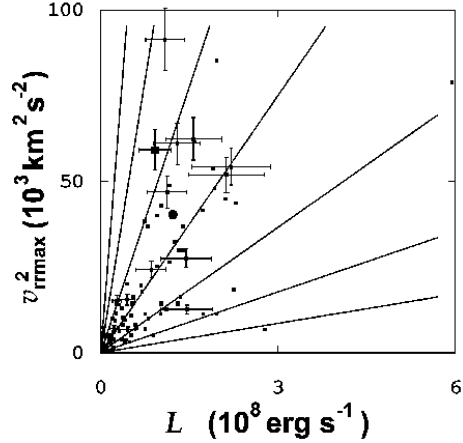}
\caption{Plots of square of the rotation velocity $ v^2_\mathrm{rrmax}$ ($10^3 \mathrm{km}^2 \, \mathrm{s}^{-2}$) at the maximum extent of the RR versus B band luminosity $L$ ($10^8\, \mathrm{erg\,s}^{-1}$) for the 95 sample galaxies.  The 15 select galaxies have error bars that show the uncertainty range in each section of the plot.  The error bars for the remaining galaxies are omitted for clarity.  The straight lines mark the lines whose characteristics are listed in Table~\ref{tab:aline}.  The large, filled circle denotes the data point for {NGC 5448}.  The large, filled square denotes the data point for {NGC 3031}.}
\label{fig:3}
\end{figure}

When the error in calculating distance is a small factor such as when using $D_\mathrm{a}$ or $D_\mathrm{tf}$, more galaxies were sampled.  The data of $ v^2_\mathrm{rrmax}$ versus $L$ of 15 select galaxies from Fig.~\ref{fig:2} and of 80 other galaxies from \citet{bege,broe,garc,guha,korn,korn2,lisz,mann,rubi85}; and \citet{swat}, for which the RR was identified, are plotted in Fig.~\ref{fig:3}.  The other sample galaxies are listed in Table~\ref{tab:other} denoted by an integer in the $a_1$ column.  Some of the galaxies had only two data points and a significant decline in $v^2$ to indicate the end of the RR.  The distribution of the data points in Fig.~\ref{fig:3} suggested a grouping of the galaxies such that the $ v^2_\mathrm{rrmax}$ versus $L$ relations are linear for each group.  This is reflected in the plotted lines in Fig.~\ref{fig:3}.  Call each group a ``classification''.  For the calculation, the ($L$,$ v^2_\mathrm{rrmax}$) $=$ (0,0) point was included in all classifications.  The relationship of the slopes of the lines is 
\begin{equation}
\log_\mathrm{10} S_\mathrm{a} = S_\mathrm{ervd} a_1 + I_\mathrm{ervd}
\label{eq:57n} \,,
\end{equation}
where $a_1$ is an integer, $S_\mathrm{ervd} = 1.3 \pm 0.2 $ and $ I_\mathrm{ervd} = 0.31 \pm 0.01 $ were obtained by a least squares fit to produce the lowest uncertainty $\sigma_\mathrm{e}$.  The $\sigma_\mathrm{e}$ was calculated as the standard deviation of the relative difference ($\delta v^2_\mathrm{rrmax}/ v^2_\mathrm{rrmax}$) between the measured parameter value and the calculated parameter value divided by the calculated parameter value for the sample galaxies.  Because the goal is to examine the slope of the lines and because the ($L$,$ v^2_\mathrm{rrmax}$) $=$ (0,0) point was included in all lines, minimization of $\sigma_\mathrm{e}$ chooses the calculated line with the closest slope to the slope of the line from (0,0) to the data point.  A $\chi^2$ type of minimization would choose a different line for some of the sample galaxies.

\begin{table*}
\begin{minipage}[]{\columnwidth}
\caption{Interger values for the non-select galaxies in the sample.}
\label{tab:other}
\centering
\begin{tabular}{lrrrrrrlrrrrrr}
\hline
Galaxy &$a_1$&$a_2$&$f_1$&$f_2$&$j_1$&$j_2$&Galaxy &$a_1$&$a_2$&$f_1$&$f_2$&$j_1$&$j_2$\\
\hline
IC 0467&3&9&&&&&NGC 4206&2&7&&&&\\
IC 2233&4&6&4&0&1&5&NGC 4216&2&8&&&&\\
NGC 0701&3&8&&&&&NGC 4222&5&10&&&&\\
NGC 0753&4&8&&&&&NGC 4237&4&6&&&&\\
NGC 0801&5&8&&&&&NGC 4359&4&3&3&1&1&5\\
NGC 0991&2&3&2&-3&1&3&NGC 4378&3&11&&&&\\
NGC 1024&4&8&&&&&NGC 4388&5&5&&&&\\
NGC 1035&4&6&&&&&NGC 4395&6&3&6&-1&-1&5\\
NGC 1042&3&4&3&-2&3&3&NGC 4448&4&7&&&&\\
NGC 1085&5&6&&&&&NGC 4605&4&10&&&&\\
NGC 1087&4&8&&&&&NGC 4647&3&7&&&&\\
NGC 1169&&&&&4&5&NGC 4654&4&1&&&&\\
NGC 1325&2&9&&&&&NGC 4682&3&7&&&&\\
NGC 1353&5&6&&&&&NGC 4689&5&9&&&&\\
NGC 1357&4&9&&&&&NGC 4698&4&6&&&&\\
NGC 1417&5&9&&&&&NGC 4772&&&&&1&3\\
NGC 1421&3&4&&&&&NGC 4845&3&7&&&&\\
NGC 1515&5&4&&&&&NGC 4866&4&9&4&7&3&4\\
NGC 1560&4&10&5&2&&&NGC 5023&4&8&4&-1&1&4\\
NGC 1620&3&5&&&&&NGC 5107&&&&&0&4\\
NGC 2715&2&7&&&&&NGC 5229&4&6&3&4&2&4\\
NGC 2742&3&5&&&&&NGC 5297&4&6&0&0&2&4\\
NGC 2770&&&&&3&4&NGC 5301&4&8&3&5&3&3\\
NGC 2844&3&13&&&&&NGC 5377&&&&&2&3\\
NGC 2903&5&11&3&5&&&NGC 5448&5&7&3&3&1&2\\
NGC 2998&3&2&&&&&NGC 5474&1&12&1&-2&0&4\\
NGC 3054&4&11&&&&&NGC 6503&5&9&5&4&&\\
NGC 3067&5&15&&&&&NGC 6814&7&6&8&-3&&\\
NGC 3109&5&5&5&2&&&NGC 7171&3&6&&&&\\
NGC 3118&4&4&4&1&2&3&NGC 7217&4&8&&&&\\
NGC 3200&4&15&&&&&NGC 7537&4&-1&&&&\\
NGC 3432&4&10&4&6&4&5&NGC 7541&4&0&&&&\\
NGC 3495&2&5&&&&&UGC 01281&4&5&4&-4&0&3\\
NGC 3593&5&11&&&&&UGC 02259&5&8&5&1&&\\
NGC 3600&4&5&4&6&4&5&UGC 03137&5&9&3&4&3&5\\
NGC 3626&&&&&4&6&UGC 03685&4&7&5&-4&0&6\\
NGC 3672&3&8&&&&&UGC 05459&5&6&4&2&2&6\\
NGC 3900&&&&&1&3&UGC 07089&4&0&4&-2&0&4\\
NGC 4010&3&8&4&3&2&5&UGC 07125&4&4&4&-1&1&3\\
NGC 4051&3&7&1&3&&&UGC 07321&2&7&4&8&6&4\\
NGC 4062&4&5&&&&&UGC 07774&4&8&4&1&2&5\\
NGC 4096&4&7&3&-2&2&4&UGC 08246&3&2&3&-2&0&5\\
NGC 4138&&&&&2&5&UGC 09242&&&&&0&23\\
NGC 4144&3&6&4&1&2&6&UGC 10205&1&10&&&&\\
NGC 4178&4&6&&&&\\
\hline
\end{tabular}
\end{minipage}
\end{table*}

\begin{table}
\begin{minipage}[]{\columnwidth}
\caption{Data for the integer values of $a_1$ of Eq.~(\ref{eq:57ads}) shown as the plotted lines in Fig.~\ref{fig:3}. }
\label{tab:aline} 
\centering
\footnotesize
\begin{tabular}{lrllll}
\hline
$a_1$\footnote{The integer denoting the exponential value in Eq.~(\ref{eq:57ads}).}
&No.\footnote{The number of data points for each $a_1$ value.}
&Corr.\footnote{The correlation coefficient rounded to two decimal places.}
&{F Test} 
&$S_\mathrm{a}$\footnote{The least squares fit slope of the $ v^2_\mathrm{rrmax}$ - $L$ lines.}\footnote{The lines were calculated using ($L$,$ v^2_\mathrm{rrmax}$) $=$ (0,0) as a data point.}
&$I_\mathrm{a}$\footnote{The least squares fit intercept of the $v^2_\mathrm{rrmax}$ - $L$ lines.} \\
\hline
1 &2&1.00&1.00&\phantom{00}2.482$\pm$0.004&\phantom{-}0.006$\pm$0.006\\
2 &7&0.97&0.93&\phantom{00}6.6\phantom{00}$\pm$0.6&\phantom{-}0.1\phantom{00}$\pm$0.7\\
3 &20&0.99&0.96&\phantom{0}12.8\phantom{00}$\pm$0.5&-0.8\phantom{00}$\pm$0.8\\
4 &39&0.98&0.92&\phantom{0}23.2\phantom{00}$\pm$0.7&\phantom{-}0.7\phantom{00}$\pm$0.7\\
5 &24&0.97&0.90&\phantom{0}42\phantom{.500}$\pm$2&\phantom{-}1\phantom{00.4}$\pm$2\\
6 &2&1.00&1.00&\phantom{0}84.3\phantom{00}$\pm$0.3&-0.2\phantom{00}$\pm$0.2\\
7 &1&1.00&1.00&261&\phantom{-}0\\
\hline
\end{tabular}
\end{minipage}
\end{table}

Therefore, 
\begin{equation}
S_\mathrm{a} = K_\mathrm{a_1} \, B_\mathrm{a_1}^{a_1} 
\label{eq:57o} \,,
\end{equation}
where $I_\mathrm{a} =0$ and $K_\mathrm{a_1} = 1.3 \pm 0.2$ and $ B_\mathrm{a_1}=2.06\pm 0.07$ at 1 $\sigma$.  The values of $K_\mathrm{a_1}$ and $B_\mathrm{a_1}$ were chosen as the maximum $B_\mathrm{a_1}$ value that yielded a correlation coefficient for each of the lines from the origin to the data points (see Table~\ref{tab:aline} and Fig.~\ref{fig:3}) greater than 0.90, that yielded a minimum $\sigma_\mathrm{e}$, and that rejected the null hypothesis of the test in Appendix~A with a confidence greater than 0.95. 

Combining Eqs.~(\ref{eq:57m}) and (\ref{eq:57o}) yields 
\begin{equation}
\frac{v^2_\mathrm{rrmax}}{10^3 \, \mathrm{km}^2 \, \mathrm{s}^{-2}} = K_\mathrm{a_1} \, B_\mathrm{a_1}^{a_1} \, \frac{L}{10^8\, \mathrm{erg\,s}^{-1}} \pm \, \sigma_\mathrm{e}
\label{eq:57ads} \,,
\end{equation}
where $\sigma_\mathrm{e} = 21\%$.  Note $\sigma_\mathrm{e} = 16\%$ for only the select galaxies.  The large, filled circle in Fig.~\ref{fig:3} denotes the data point for {NGC 5448} \mbox{($\delta v^2_\mathrm{rrmax} / v^2_\mathrm{rrmax} = -0.35$)}.  The large, filled square in Fig.~\ref{fig:3} denotes the data point for {NGC 3031} \mbox{($\delta v^2_\mathrm{rrmax} / v^2_\mathrm{rrmax} = 0.25$)}. 

Tables~\ref{tab:other} and \ref{tab:integer1} lists the $a_1$ values for the sample galaxies.  Table~\ref{tab:constants} lists the minimum correlation coefficient $C_\mathrm{cmin}$ of the lines, the constants $K_\mathrm{x}$, and the exponential bases $B_\mathrm{x}$, where the subscript ``x'' denotes the generic term.  

\begin{table*}
\begin{minipage}[]{140mm}
\caption{First approximation integer values for the select galaxies. }
\label{tab:integer1} 
\centering
\begin{tabular}{lrrrrrrrrrrrrrrrrrrrrrrrr}
\hline
Galaxy&$a_1$&$b_1$&$c_1$&$d_1$&$e_1$&$f_1$&$g_1$&$h_1$&$i_1$&$j_1$&$k_1$&$l_1$\\
\hline
{NGC 0224}&5&5&5&3&4&4&3&3&4&&&4\\
{NGC 0598}&4&6&3&6&6&5&5&3&6&&&6\\
{NGC 3031}&5&5&5&5&4&3&3&2&4&3&5&7\\
{NGC 0300}&5&7&3&6&7&5&5&2&7&&&7\\
{NGC 2403}&5&6&4&6&6&4&4&2&6&1&4&6\\
{NGC 5457}&4&5&5&1&2&1&4&3&1&4&5&4\\
{NGC 4258}&3&2&1&4&5&2&2&2&3&2&1&2\\
{NGC 0925}&5&6&4&4&2&2&4&2&4&2&3&6\\
{NGC 2841}&&1&&&&4&4&4&3&2&5&1\\
{NGC 3198}&4&6&5&2&1&2&4&3&3&2&6&5\\
{NGC 4414}&5&4&4&4&4&2&3&2&3&1&2&6\\
{NGC 3319}&3&6&4&2&3&3&4&2&5&1&5&4\\
{NGC 7331}&5&3&3&5&5&3&3&3&4&3&3&4\\
{NGC 4535}&4&4&3&3&4&2&1&1&5&3&4&3\\
{NGC 4321}&4&4&4&2&2&2&1&2&4&3&2&3\\
{NGC 4548}&6&1&2&10&8&4&2&2&6&9&16&7\\
\hline
\end{tabular}
\end{minipage}
\end{table*}

The average difference between $L$ and the luminosity calculated using $D_\mathrm{tf}$ is 0.38$L$ for the select galaxies.  The relative difference in $ v^2_\mathrm{rrmax}$ included the measurement uncertainty and the uncertainty that the RR may extend farther than the measured point such as seen for {NGC 4258}.  The relative differences $\delta L / L = 0.3$ and $\delta v^2_\mathrm{rrmax} / v^2_\mathrm{rrmax} = 0.1$ are shown as error bars in Fig.~\ref{fig:3} for the select galaxies.  

Other parametric relationships to $L$ in the RR were calculated using the same procedure that was used to evaluate the $ v^2_\mathrm{rrmax}$ - $L$ relation.  Because these equations involved $ R_\mathrm{rrmax}$, the calculation considered only the 15 select galaxies.  The resulting equations are:
\begin{equation}
\frac{R_\mathrm{rrmax}}{\mathrm{kpc}} = K_\mathrm{b_1} \, B_\mathrm{b_1}^{b_1} \, \frac{L}{10^8\, \mathrm{erg\,s}^{-1}}\,\pm 14\% 
\label{eq:57bds} \,;
\end{equation}

\begin{equation}
\frac{ R_\mathrm{rrmax} v^2_\mathrm{rrmax}}{10^3 \, \mathrm{kpc} \, \mathrm{km}^2 \, \mathrm{s}^{-2}} = K_\mathrm{c_1} \, B_\mathrm{c_1}^{c_1} \, \frac{L}{10^8\, \mathrm{erg\,s}^{-1}}\,\pm 15\% 
\label{eq:57cds} \,;
\end{equation}

\begin{equation}
\frac{  v^2_\mathrm{rrmax}/ R_\mathrm{rrmax} }{10^3 \, \mathrm{kpc}^{-1} \, \mathrm{km}^2 \, \mathrm{s}^{-2}} = K_\mathrm{d_1} \, B_\mathrm{d_1}^{d_1} \, \frac{L}{10^8\, \mathrm{erg\,s}^{-1}}\,\pm 11\% 
\label{eq:57dds} \,;\, \, \mathrm{and}
\end{equation}

\begin{equation}
\frac{S_\mathrm{rr}}{ 10^3 \, \mathrm{kpc}^{-1} \, \mathrm{km}^2 \, \mathrm{s}^{-2}} = K_\mathrm{e_1} \, B_\mathrm{e_1}^{e_1} \, \frac{L}{10^8\, \mathrm{erg\,s}^{-1}}\, \pm 13\% 
\label{eq:57eds} \,;
\end{equation}
where first approximation integer values are listed in Table~\ref{tab:integer1} and the $C_\mathrm{cmin}$, $K_\mathrm{x}$, and $B_\mathrm{x}$ are listed in Table~\ref{tab:constants}.

\begin{table}
\begin{minipage}[]{\columnwidth}
\caption{Values of the minimum correlation coefficients $C_\mathrm{cmin}$, constants $K_\mathrm{x}$, and exponent bases $B_\mathrm{x}$ for the first approximation equations. }
\label{tab:constants} 
\centering
\footnotesize
\begin{tabular}{lllll}
\hline
Integer\footnote{The integer of the exponential value of parametric relationship that denotes the applicable equation.}
&$C_\mathrm{cmin}$\footnote{The minimum correlation coefficient rounded to two decimal places of the lines of the parametric relationship.}
&$K_\mathrm{x}$\footnote{The constant of proportionality at 1$\sigma$ of the parametric relationship.} 
&$B_\mathrm{x}$\footnote{The exponent base at 1$\sigma$ of the parametric relationship.}\\
\hline
$a_1$ &0.97&\phantom{01}1.3\phantom{00}$\pm$\phantom{1}0.2&2.06$\pm$0.07\\
$b_1$ &0.86&\phantom{01}0.34\phantom{0}$\pm$\phantom{1}0.04&1.88$\pm$0.05\\
$c_1$ &0.96&\phantom{1}19\phantom{.000}$\pm$\phantom{0}1&1.89$\pm$0.05\\
$d_1$ &0.99&\phantom{01}1.0\phantom{00}$\pm$\phantom{1}0.1&1.57$\pm$0.06\\
$e_1$ &0.97&\phantom{01}0.49\phantom{0}$\pm$\phantom{1}0.06&1.71$\pm$0.04\\
$f_1$ &0.96&\phantom{01}8.6\phantom{00}$\pm$\phantom{1}0.7&1.57$\pm$0.04\\
$g_1$ &0.93&\phantom{01}4.4\phantom{00}$\pm$\phantom{1}0.4&1.77$\pm$0.06\\
$h_1$ &0.97&150\phantom{.000}$\pm$10&2.5\phantom{0}$\pm$0.1\\
$i_1$ &0.95&\phantom{01}0.09\phantom{0}$\pm$\phantom{1}0.02&1.9\phantom{0}$\pm$0.2\\
$l_1$ &0.96&\phantom{01}0.075$\pm$\phantom{1}0.009&1.95$\pm$0.02\\
\hline
\end{tabular}
\end{minipage}
\end{table}

Equations~(\ref{eq:57ads})--(\ref{eq:57dds}) suggest at least two of the four integers $a_1$, $b_1$, $c_1$, and $d_1$ are functions of the other two.

The end of the EOR was considered to be at the largest radius $R_\mathrm{eormax}$ (kpc) along the major axis that H\,{\sc i} is in orbit around the galaxy.  Several galaxies have a H\,{\sc i} measured point on one side of the galaxy at a greater $R_\mathrm{major}$ than the other side.  Because averaging the $v$ from each side of the galaxy was required to reduce the $\vert \vec{K} \bullet \vec{a}_\mathrm{o} \vert $ effect that causes non-circular rotation, the outermost measured point used in the calculations was at the $R_\mathrm{eormax}$ with data points on both sides of the galaxy.  At $R > R_\mathrm{eormax}$ particles are no longer in orbit.  For particles other than H\,{\sc i}, the $m_\mathrm{s}/ m_\mathrm{\iota}$ is less.  Therefore, their maximum radius in a galaxy is less.  {NGC 4321} has H\,{\sc i} extending farther than a radius of four arcminutes.  However, \citet{knap} concluded the H\,{\sc i} motion beyond four arcminutes is not disk rotation.  Therefore, $R_\mathrm{eormax}$ was considered to be at four arcminutes for {NGC 4321}. 

Because $\epsilon \propto M_\mathrm{t \epsilon}$ and $\epsilon \propto M_\Delta$, $M_\mathrm{eormax} \propto \epsilon \propto L $, where $M_\mathrm{eormax}$ is the mass within $R_\mathrm{eormax}$ of Eq.~(\ref{eq:51c}).  The rotation velocity $v_\mathrm{eormax}$ of the galaxy at a radius of $R_\mathrm{eormax}$ can be used to compare parameters among galaxies.  

Parameter relations to $L$ in the EOR were calculated using the same procedure used to evaluate the $ v^2_\mathrm{rrmax}$ -- $L$ relation. 

\begin{figure}
\centering 
\includegraphics[width=0.5\textwidth]{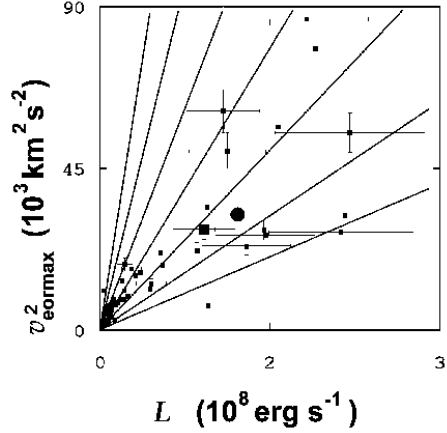}
\caption{Plots of the square of the H\,{\sc i} rotation velocity $ v^2_\mathrm{eormax}$ ($10^3 \, \mathrm{km}^2 \, \mathrm{s}^{-2}$) at extreme outer region versus B band luminosity $L$ ($10^8\, \mathrm{erg\,s}^{-1}$) for the 50 sample galaxies.  The 16 select galaxies have error bars that show the uncertainty level in each section of the plot.  The error bars for the remaining galaxies are omitted for clarity.  The large, filled circle denotes the data point for {NGC 5448}.  The large, filled square denotes the data point for {NGC 3031}.}
\label{fig:4}
\end{figure}

The data of $ v^2_\mathrm{eormax}$ versus $L$ of 16 select galaxies and of 34 other galaxies from \citet{bege,broe,garc,korn,korn2,lisz,mann}; and \citet{swat} for a total of 50 sample galaxies are plotted in Fig.~\ref{fig:4}.  The 34 other galaxies are listed in Table~\ref{tab:other} denoted by an integer in the $f_1$ column.  The uncertainty in determining the value of the outermost point of the EOR is greater than determining the end of the RR.  Because the determination of the outermost value of the parameters requires instrumentation sensitive enough to measure the values, data prior to 1990 of 46 other galaxies used in the $v^2_\mathrm{rrmax}$ --$L$ relation were omitted.  The equations involving $R_\mathrm{eormax}$ used only the 16 select galaxies.

The resulting equations are:
\begin{equation}
\frac{v^2_\mathrm{eormax}}{10^3 \, \mathrm{km}^2 \, \mathrm{s}^{-2}} = K_\mathrm{f_1} \, B_\mathrm{f_1}^{f_1} \, \frac{L}{10^8\, \mathrm{erg\,s}^{-1}} \pm \, 13\% 
\label{eq:57gds} \,,
\end{equation}
where the large, filled circle in Fig.~\ref{fig:4} denotes the data point for {NGC 5448} \mbox{($\delta v^2_\mathrm{ eormax } / v^2_\mathrm{ eormax } = -0.20$)} and the large, filled square denotes the data point for {NGC 3031} \mbox{($\delta v^2_\mathrm{ eormax } / v^2_\mathrm{ eormax } = -0.08$)};
\begin{equation}
\frac{R_\mathrm{eormax}}{\mathrm{kpc}} = K_\mathrm{g_1} \, B_\mathrm{g_1}^{g_1} \, \frac{L}{10^8\, \mathrm{erg\,s}^{-1}}\, \pm \,16\% 
\label{eq:57fds} \,;
\end{equation}
\begin{equation}
\frac{ R_\mathrm{eormax} v^2_\mathrm{eormax}}{10^3 \, \mathrm{kpc} \, \mathrm{km}^2 \, \mathrm{s}^{-2}} = K_\mathrm{h_1} \, B_\mathrm{h_1}^{h_1} \, \frac{L}{10^8\, \mathrm{erg\,s}^{-1}}\, \pm \, 19\% 
\label{eq:57hds} \,; \, \, \mathrm{and}
\end{equation}
\begin{equation}
\frac{ v^2_\mathrm{eormax}/ R_\mathrm{eormax} }{10^3 \, \mathrm{kpc}^{-1} \, \mathrm{km}^2 \, \mathrm{s}^{-2}} = K_\mathrm{i_1} \, B_\mathrm{i_1}^{i_1} \, \frac{L}{10^8\, \mathrm{erg\,s}^{-1}}\, \pm \, 15\% 
\label{eq:57ids} \,;
\end{equation}
where first approximation integer values are listed in Tables~\ref{tab:other} and \ref{tab:integer1} and the $C_\mathrm{cmin}$, $K_\mathrm{x}$, and $B_\mathrm{x}$ are listed in Table~\ref{tab:constants}.

The uncertainty in calculating $\vert \vec{K} \bullet \vec{a}_\mathrm{o} \vert $ depends on the accuracy of measuring the distance $D$ (Mpc) to neighbor galaxies that are without a $D_\mathrm{a}$ and without a $D_\mathrm{tf}$ measurement.  The cell structure of galaxy clusters \citep{aaro2,cecc,hodg,huds,lilj,rejk} suggests a systematic error in $z$ relative to $D$ of a galaxy.  Therefore, using $z$ and the Hubble Law to determine $D$ introduces a large error.  However, the cell model also suggests neighboring galaxies have a similar $D/z$ ratio in the first approximation.  For those few galaxies near our line of sight to the sample galaxy and near the sample galaxy, \citet{hodg} found a $z$ change caused by the light passing close to galaxies.  For a single galaxy this effect is small and was ignored. Therefore,
\begin{equation}
D_i = \frac{z_i}{z_p} D_p
\label{eq:or2} \,,
\end{equation}
where $z_p$ and $ D_p$ (Mpc) are the redshift and distance of the sample galaxy, respectively, and $z_i$ and $ D_i$ (Mpc) are the redshift and calculated distance of the $i^{th}$ neighbor galaxy, respectively.  The $ D_p =D_\mathrm{a}$ for the select galaxies and $ D_p =D_\mathrm{tf}$ for the other sample galaxies.  The $L$ of the $i^{th}$ neighbor galaxy was calculated using $ D_i$, $m_\mathrm{b}$, and $E_{\mathrm{xt}}$. 

Because larger $\vert z_i - z_p \vert$ implies larger error in $D_i$, the $N_\mathrm{sources}$ and $N_\mathrm{sinks}$ of Eq.~(\ref{eq:51ao}) was limited to the number $N$ of galaxies with the largest influence on $\vert \vec{\nabla} \rho \vert$ of the sample galaxy, where the $\vec{\nabla} \rho$ of Eq.~(\ref{eq:51ao}) is evaluated at the center of the sample galaxy.  The $N=7$ was chosen because it produced the highest correlation in the calculations.  Also, $6 \leq N \leq 9$ produced acceptable correlation in the calculations.

The calculation of $\vert \vec{K} \bullet \vec{a}_\mathrm{o} \vert $ also requires knowledge of the orientation of the sample galaxy.  The direction of the sample galaxy's polar unit vector $\vec{e}_\mathrm{polar}$ was defined as northward from the center of the sample galaxy along the polar axis of the galaxy.  \citet{hu,pere}; and \citet{truj} found an alignment of the polar axis of neighboring spiral galaxies.  Therefore, the orientation of $\vec{e}_\mathrm{polar}$ with the higher $\vert \vec{e}_\mathrm{polar} \bullet \vec{a}_\mathrm{o} \vert$ was chosen for the calculations.

The major axis unit vector $\vec{e}_\mathrm{major}$ was defined as eastward from the center of the sample galaxy along the major axis of the galaxy.  The minor axis unit vector \mbox{$\vec{e}_\mathrm{minor} \equiv \vec{e}_\mathrm{major} \times \vec{e}_\mathrm{polar}$}. 

At equal $R_\mathrm{major}>R_\mathrm{rrmax}$ on opposite sides of the galaxy, the $L$ terms of Eq.~(\ref{eq:51c}) are equal and the $M$ terms are nearly equal.  Define asymmetry $A_\mathrm{sym}$, 
\begin{equation}
A_\mathrm{sym} \equiv (v_\mathrm{h}^2 - v_\mathrm{l}^2) \vert _{R} 
\label{eq:or4} \,,
\end{equation}
where $v_\mathrm{h}$ and $v_\mathrm{l}$ are the larger $v$ and smaller $v$ at the same $R_\mathrm{major}$ ($\vert_{R}$) on opposite sides of the galaxy, respectively.

The $A_\mathrm{sym}$ is a function of $R_\mathrm{major}$.  The $A_\mathrm{sym}$ and $ \vert \vec{K} \bullet \vec{a}_\mathrm{o} \vert$ are sensitive to both radial (non-circular) motion of particles and to variation of $v$ due to torque.  For the first approximation the maximum asymmetry $A_\mathrm{symax}$ in the OR depends predominately on $ \vert \vec{K} \bullet \vec{a}_\mathrm{o} \vert$.  Because $A_\mathrm{symax}$ is minimally dependant on $R$, $A_\mathrm{symax}$ is comparable among galaxies.

\begin{figure}
\centering 
\includegraphics[width=0.5\textwidth]{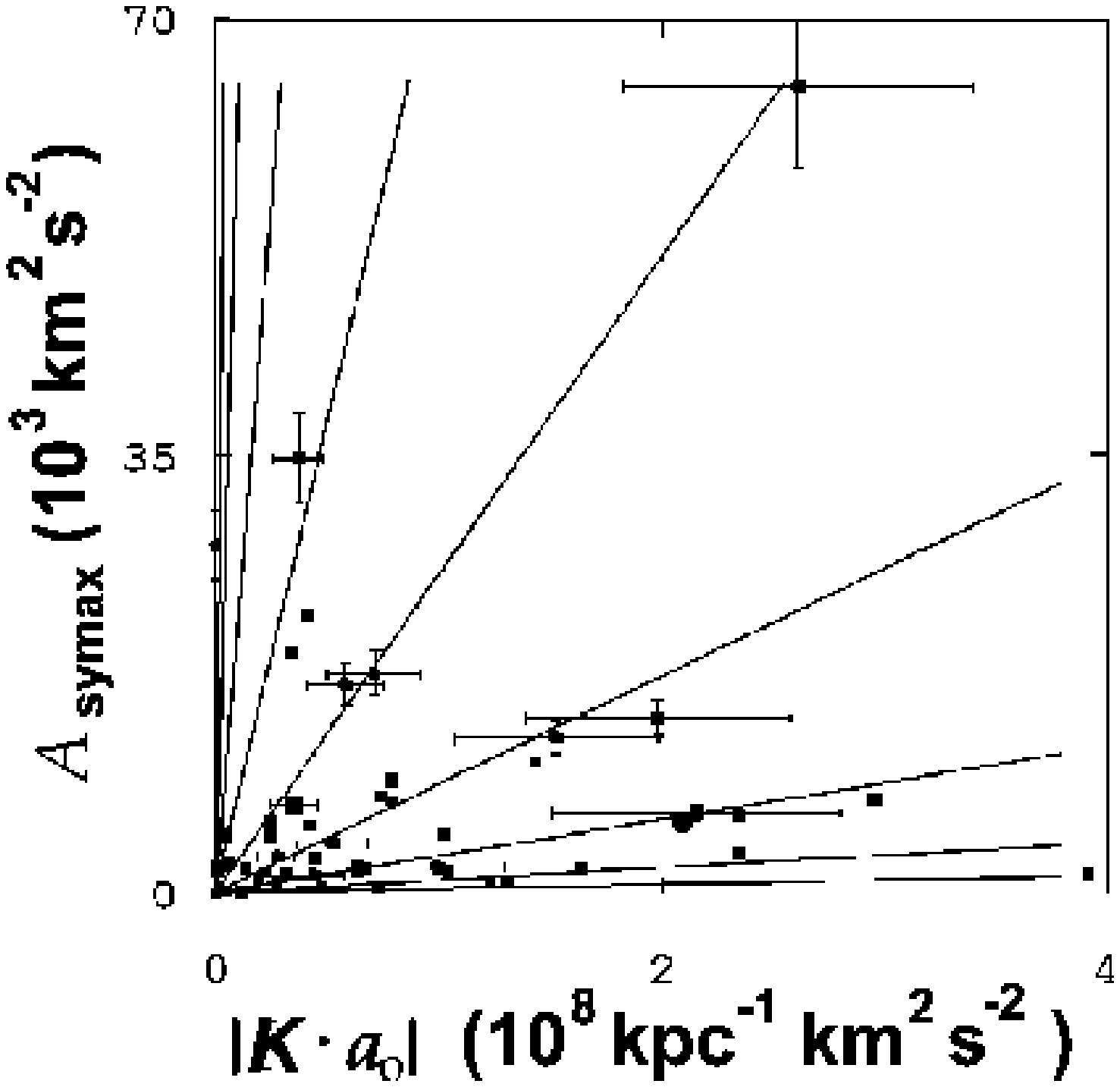}
\caption{Plots of maximum asymmetry $ A_\mathrm{symax} $ ($10^3 \, \mathrm{km}^2 \, \mathrm{s}^{-2}$) versus $\vert \vec{K} \bullet \vec{a}_\mathrm{o} \vert $ ($10^3 \, \mathrm{kpc}^{-1} \, \mathrm{km}^2 \, \mathrm{s}^{-2}$) for the 50 sample galaxies.  The 13 select galaxies have error bars that show the uncertainty level in each section of the plot.  .  The error bars for the remaining galaxies are omitted for clarity.  The large, filled circle denotes the data point for {NGC 5448}.  The large, filled square denotes the data point for {NGC 3031}.}
\label{fig:5}
\end{figure}

{NGC 0224}, {NGC 0300}, and {NGC 0598} were omitted from the sample because their $z < 0$ and their neighbor galaxies may have positive or negative $z$.  Therefore, evaluating the distance of neighbor galaxies of these galaxies was considered unreliable.  Table~\ref{tab:ORdata} lists the $ \vert \vec{K} \bullet \vec{a}_\mathrm{o} \vert$, $A_\mathrm{symax}$, and $A_\mathrm{symax}$ references for the remaining 13 select galaxies.  The sample consisted of 13 select galaxies and of 36 other galaxies from \citet{garc,korn,korn2}; and \citet{swat} for a total of 49 sample galaxies.  The 36 other galaxies are listed in Table~\ref{tab:other} denoted by an integer in the $j_1$ column.  The plot of $A_\mathrm{symax}$ versus $\vert \vec{K} \bullet \vec{a}_\mathrm{o} \vert$ is shown in Fig.~\ref{fig:5}. 

The same procedure that was used to evaluate the $ v^2_\mathrm{rrmax}$ -- $L$ relation was used to evaluate the $A_\mathrm{symax}$ -- $\vert \vec{K} \bullet \vec{a}_\mathrm{o} \vert$ relation.  The result is
\begin{equation}
\frac{A_\mathrm{symax}}{10^3 \, \mathrm{km}^2 \, \mathrm{s}^{-2}} = K_\mathrm{j_1} B_\mathrm{j_1}^{j_1} \, \frac{\vert \vec{K} \bullet \vec{a}_\mathrm{o} \vert} {10^3 \, \mathrm{kpc}^{-1} \, \mathrm{km}^2 \, \mathrm{s}^{-2}} \pm \, 28\% 
\label{eq:or4a} \,,
\end{equation}
where first approximation integer values are listed in Tables~\ref{tab:other} and \ref{tab:integer1}, $C_\mathrm{cmin}=0.98$, $K_\mathrm{j_1} = (1.000 \pm 0.001)$, $B_\mathrm{j_1}=2.94 \pm 0.09$, and 
\begin{eqnarray}
\vec{K} & = & \phantom{-} (3.60\times10^{-6}) \, \vec{e}_\mathrm{polar}  \nonumber \\*
& & -(6.81\times10^{-6}) \, \vec{e}_\mathrm{major} \nonumber \\*
& & -(1.18\times10^{-5}) \, \vec{e}_\mathrm{minor} \, \mathrm{km \, kpc}^{-1}
\end{eqnarray}
at 1$\sigma$.  The value of $\vec{K} $ was chosen such that \mbox{$ K_\mathrm{j_1} \approx 1.00$}. 

\begingroup
\begin{table}
\begin{minipage}[]{\columnwidth}
\caption{Asymmetry and $ \vert \vec{K} \bullet \vec{a}_\mathrm{o} \vert$ data for the select galaxies.}
\label{tab:ORdata}
\centering
\begin{tabular}{lrlc}
\hline
Galaxy &
$A_\mathrm{symax}$\footnote{The units are $10^3 \, \mathrm{km}^2 \, \mathrm{s}^{-2}$.  The uncertainty is $\pm 10\%$.}&$ \vert \vec{K} \bullet \vec{a}_\mathrm{o} \vert$\footnote{ The units are $10^3 \, \mathrm{kpc}^{-1} \, \mathrm{km}^2 \, \mathrm{s}^{-2}$.  The uncertainty is $\pm 30\%$.}&
Ref.\footnote{ References: [1]\citet{krum}, [2]\citet{bege}, [3] \citet{fill}, [4]\citet{rots}, [5]\citet{vana}, [6]\citet{bosm3}, [7]\citet{moor}, [8]\citet{vana2}, [9]\citet{semp}, [10]\citet{brai}, [11]\citet{chin}, [12]\citet{voll}, and [13]\citet{bosm2}.}\\
\hline
{NGC 0925} &12.5&1.53&[1]\\
{NGC 2403} &2.0&0.99&[2]\\
{NGC 2841} &4.0&0.52&[3][4]\\
{NGC 3031} &7.1&0.35&[4]\\
{NGC 3198} &2.9&0.27&[5][6]\\
{NGC 3319} &1.4&0.44&[7]\\
{NGC 4258} &14.0&2.0&[8]\\
{NGC 4321} &64.7&2.6&[9]\\
{NGC 4414} &6.5&2.2&[10]\\
{NGC 4535} &16.7&0.58&[11]\\
{NGC 4548} &27.9&0.0018&[12]\\
{NGC 5457} &34.8&0.37&[13]\\
{NGC 7331} &17.6&0.71&[6]\\
\hline
\end{tabular}
\end{minipage}
\end{table}
\endgroup

The large, filled circle in Fig.~\ref{fig:5} denotes the data point for {NGC 5448} \mbox{($\delta A_\mathrm{symax}/A_\mathrm{symax}=-0.08$)}.  The large, filled square denotes the data point for {NGC 3031} \mbox{($\delta A_\mathrm{symax}/A_\mathrm{symax}=-0.21$)}.  For the 13 select galaxies, $\sigma_\mathrm{e}= 17\%$ [excluding the \mbox{$(A_\mathrm{symax},\vert\vec{K}\bullet\vec{a}_\mathrm{o}\vert)=(0,0)$} point].

Because the net effect of $F_\mathrm{s}$ and $F_\mathrm{g} $ from the host galaxy is weak at the outer edge of the galaxy, the slope $S_\mathrm{eor}$ ($\mathrm{kpc}^{-1} \, \mathrm{km}^2 \, \mathrm{s}^{-2}$) of the H\,{\sc i} RC for the end of the EOR was compared among the 13 select galaxies.  The $S_\mathrm{eor}$ were found by fitting a least squares straight line to the outermost two or more data points of the RC with a 0.97 or higher correlation coefficient.  The same procedure that was used to evaluate the $ v^2_\mathrm{rrmax}$ -- $L$ relation was used to evaluate the $S_\mathrm{eor}$ -- $\vert \vec{K} \bullet \vec{a}_\mathrm{o} \vert$ relation.  The result is
\begin{eqnarray}
\frac{S_\mathrm{eor}}{ 10^3 \mathrm{kpc}^{-1} \, \mathrm{km}^2 \, \mathrm{s}^{-2}} & = & K_\mathrm{k_1} B_\mathrm{k_1}^{k_1} \, \frac{\vert \vec{K} \bullet \vec{a}_\mathrm{o} \vert} { 10^3 \mathrm{kpc}^{-1} \, \mathrm{km}^2 \, \mathrm{s}^{-2}} \nonumber \\*
& & + I_\mathrm{seor}  \pm \, 15\% 
\label{eq:or1a} \,,
\end{eqnarray}
where $C_\mathrm{cmin}=0.96$, $ K_\mathrm{k_1} = 0.64 \pm 0.09$, $B_\mathrm{k_1} = 1.74 \pm 0.06$, and $ I_\mathrm{seor} = -4.8\pm0.1$ at 1$\sigma$.  Tables~\ref{tab:integer1} lists the $k_1$ values for the 13 select galaxies.

As seen in Fig.~\ref{fig:2} the measured points in the OR appear nearly linear rather than a smoothly varying curve in most of the select galaxies.  Therefore, the linear approximation of the RC slope in the OR was considered more appropriate than a smoothly varying function.  Define $S_\mathrm{or}$ ($\mathrm{km}^2 \, \mathrm{s}^{-2} \, \mathrm{kpc}^{-1}$) as the slope from the $R_\mathrm{rrmax}$ data point to the beginning of the $S_\mathrm{eor}$ of the H\,{\sc i} RC.  For the 15 select galaxies excluding {NGC 2841}, following the same procedure as for finding Eq.~(\ref{eq:57ads}) yields
\begin{equation}
\frac{S_\mathrm{or}}{10^3 \, \mathrm{km}^2 \, \mathrm{s}^{-2} \, \mathrm{kpc}^{-1}} = - \, K_\mathrm{l_1} \, B_\mathrm{l_1}^{l_1} \, \frac{L}{10^8\, \mathrm{erg\,s}^{-1}} + I_\mathrm{or} \pm \, 15\% 
\label{eq:ksor} \,,
\end{equation}
where the integer values are listed in Table~\ref{tab:integer1}.  The $C_\mathrm{cmin}$, $K_\mathrm{l_1}$, and $B_\mathrm{l_1}$ values are listed in Table~\ref{tab:constants} and $ I_\mathrm{or}=1.6 \pm 0.1$. 

\subsection{Second approximation}

Equation~(\ref{eq:51c}) indicates the parameters of galaxies depend on $L$ and \mbox{$\vert \vec{K} \bullet \vec{a}_\mathrm{o} \vert $}.  The first approximation selected either $L$ or $\vert \vec{K} \bullet \vec{a}_\mathrm{o} \vert $ depending upon which is predominant.  The second approximation considered the less predominant term to be a systematic correction term.  Posit the residual between the measured parameter value and the calculated first approximation parameter value is proportional to the less predominant term.  Appling the same procedure as for finding Eq.~(\ref{eq:57ads}) yields: 

\begin{eqnarray}
\frac{v^2_\mathrm{rrmax}}{10^3 \, \mathrm{km}^2 \, \mathrm{s}^{-2}} = K_\mathrm{a_1} \, B_\mathrm{a_1}^{a_1} \, \frac{L}{10^8\, \mathrm{erg\,s}^{-1}} + (-1)^{s_a} K_\mathrm{a_2} B_\mathrm{a_2}^{a_2} \frac{\vert \vec{K} \bullet \vec{a}_\mathrm{o} \vert} {10^3 \,  \mathrm{kpc}^{-1} \, \mathrm{km}^2 \, \mathrm{s}^{-2}} \pm 10\% 
\label{eq:add1c} \,,
\end{eqnarray}
where $ K_{a_2} =0.15 \pm 0.02$, $ B_\mathrm{a_2}=2.00 \pm 0.03$ at 1$\sigma$, for {NGC 5448} \mbox{($\delta v^2_\mathrm{rrmax} / v^2_\mathrm{rrmax} = -0.04$)}, and for {NGC 3031} \mbox{($\delta v^2_\mathrm{rrmax} / v^2_\mathrm{rrmax} = -0.03$)};
\begin{eqnarray}
\frac{R_\mathrm{rrmax}}{\mathrm{kpc}} = K_\mathrm{b_1} \, B_\mathrm{b_1}^{b_1} \, \frac{L}{10^8\, \mathrm{erg\,s}^{-1}} + (-1)^{s_b} K_\mathrm{b_2} B_\mathrm{b_2}^{b_2} \frac{\vert \vec{K} \bullet \vec{a}_\mathrm{o} \vert} {10^3 \,  \mathrm{kpc}^{-1} \, \mathrm{km}^2 \, \mathrm{s}^{-2}} \pm 2\% 
\label{eq:add2c} \,,
\end{eqnarray}
where $ K_\mathrm{b_2} =0.08 \pm 0.01$ and $ B_\mathrm{b_2}=1.75 \pm 0.08$ at 1$\sigma$;
\begin{eqnarray}
\frac{ R_\mathrm{rrmax} v^2_\mathrm{rrmax}}{10^3 \, \mathrm{kpc} \, \mathrm{km}^2 \, \mathrm{s}^{-2}} = K_\mathrm{c_1} \, B_\mathrm{c_1}^{c_1} \, \frac{L}{10^8\, \mathrm{erg\,s}^{-1}} + (-1)^{s_{c}} K_\mathrm{c_2} B_\mathrm{c_2}^{c_2} \frac{\vert \vec{K} \bullet \vec{a}_\mathrm{o} \vert} {10^3 \, \mathrm{kpc}^{-1} \, \mathrm{km}^2 \, \mathrm{s}^{-2}} \pm 3\% 
\label{eq:add3c} \,,
\end{eqnarray}
where $ K_\mathrm{c_2} =0.63 \pm 0.07$ and $ B_\mathrm{c_2}=1.98 \pm 0.03$ at 1$\sigma$;
\begin{eqnarray}
\frac{ v^2_\mathrm{rrmax}/ R_\mathrm{rrmax} }{10^3 \, \mathrm{kpc}^{-1} \, \mathrm{km}^2 \, \mathrm{s}^{-2}} =  K_\mathrm{d_1 } \, B_\mathrm{d_1}^{d_1} \, \frac{L}{10^8\, \mathrm{erg\,s}^{-1}} +  (-1)^{s_{d}} K_\mathrm{d_2} B_\mathrm{d_2}^{d_2} \frac{\vert \vec{K} \bullet \vec{a}_\mathrm{o} \vert} {10^3 \,  \mathrm{kpc}^{-1} \, \mathrm{km}^2 \, \mathrm{s}^{-2}} \pm 3\% 
\label{eq:add4c} \,,
\end{eqnarray}
where $ K_\mathrm{d_2} =0.024 \pm 0.007$ and $ B_\mathrm{d_2}=2.2 \pm 0.3$ at 1$\sigma$;
\begin{eqnarray}
\frac{S_\mathrm{rr}}{ 10^3 \, \mathrm{kpc}^{-1} \, \mathrm{km}^2 \, \mathrm{s}^{-2}} = K_\mathrm{e_1} \, B_\mathrm{e_1}^{e_1} \, \frac{L}{10^8\, \mathrm{erg\,s}^{-1}} + (-1)^{s_{e}} K_\mathrm{e_2} B_\mathrm{e_2}^{e_2} \frac{\vert \vec{K} \bullet \vec{a}_\mathrm{o} \vert} {10^3 \,  \mathrm{kpc}^{-1} \, \mathrm{km}^2 \, \mathrm{s}^{-2}} \pm 2\% 
\label{eq:add5c} \,,
\end{eqnarray}
where $ K_\mathrm{e_2} =0.021 \pm 0.003$ and $ B_\mathrm{e_2}=1.72 \pm 0.03$ at 1$\sigma$;
\begin{eqnarray}
\frac{v^2_\mathrm{eormax}}{10^3 \, \mathrm{km}^2 \, \mathrm{s}^{-2}} = K_\mathrm{f_1} \, B_\mathrm{f_1}^{f_1} \, \frac{L}{10^8\, \mathrm{erg\,s}^{-1}} + (-1)^{s_{f}} K_\mathrm{f_2} B_\mathrm{f_2}^{f_2} \frac{\vert \vec{K} \bullet \vec{a}_\mathrm{o} \vert} {10^3 \,  \mathrm{kpc}^{-1} \, \mathrm{km}^2 \, \mathrm{s}^{-2}} \pm 2\% 
\label{eq:57gdsc} \,,
\end{eqnarray}
where $ K_\mathrm{f_2} =0.7 \pm 0.1$ and $ B_\mathrm{f_2}=1.9 \pm 0.1$ at 1$\sigma$, for {NGC 5448} \mbox{($\delta v^2_\mathrm{eormax} / v^2_\mathrm{eormax} = -0.03$)}, and for {NGC 3031} \mbox{($\delta v^2_\mathrm{eormax} / v^2_\mathrm{eormax} = -0.02$)};
\begin{eqnarray}
\frac{R_\mathrm{eormax}}{\mathrm{kpc}} = K_\mathrm{g_1} \, B_\mathrm{g_1}^{g_1} \, \frac{L}{10^8\, \mathrm{erg\,s}^{-1}} + (-1)^{s_{g}} K_\mathrm{g_2} B_\mathrm{g_2}^{g_2} \frac{\vert \vec{K} \bullet \vec{a}_\mathrm{o} \vert} {10^3 \,  \mathrm{kpc}^{-1} \, \mathrm{km}^2 \, \mathrm{s}^{-2}} \pm 3\% 
\label{eq:57fdsc} \,,
\end{eqnarray}
where $ K_\mathrm{g_2} =0.008 \pm 0.002$ and $ B_\mathrm{g_2}=2.05 \pm 0.09$ at 1$\sigma$;
\begin{eqnarray}
\frac{ R_\mathrm{eormax} v^2_\mathrm{eormax}}{10^3 \, \mathrm{kpc} \, \mathrm{km}^2 \, \mathrm{s}^{-2}} = K_\mathrm{h_1} \, B_\mathrm{h_1}^{h_1} \, \frac{L}{10^8\, \mathrm{erg\,s}^{-1}} + (-1)^{s_{h}} K_\mathrm{h_2} B_\mathrm{h_2}^{h_2} \frac{\vert \vec{K} \bullet \vec{a}_\mathrm{o} \vert} {10^3 \,  \mathrm{kpc}^{-1} \, \mathrm{km}^2 \, \mathrm{s}^{-2}} \pm 3\% 
\label{eq:57hdsc} \,,
\end{eqnarray}
where $ K_\mathrm{h_2} =7 \pm 2$ and $ B_\mathrm{h_2}=2.06 \pm 0.08$ at 1$\sigma$;
\begin{eqnarray}
\frac{ v^2_\mathrm{eormax}/ R_\mathrm{eormax} }{10^3 \, \mathrm{kpc}^{-1} \, \mathrm{km}^2 \, \mathrm{s}^{-2}} = K_\mathrm{i_1} \, B_\mathrm{i_1}^{i_1} \, \frac{L}{10^8\, \mathrm{erg\,s}^{-1}} + (-1)^{s_{i}} K_\mathrm{i_2} B_\mathrm{i_2}^{i_2} \frac{\vert \vec{K} \bullet \vec{a}_\mathrm{o} \vert} {10^3 \,  \mathrm{kpc}^{-1} \, \mathrm{km}^2 \, \mathrm{s}^{-2}} \pm 2\% 
\label{eq:57idsc} \,,
\end{eqnarray}
where $ K_\mathrm{i_2} =0.024 \pm 0.002$ and $ B_\mathrm{i_2}=1.78 \pm 0.04$ at 1$\sigma$;
\begin{eqnarray}
\frac{A_\mathrm{symax}}{10^3 \, \mathrm{km}^2 \, \mathrm{s}^{-2}} = K_\mathrm{j_1} B_\mathrm{j_1}^{j_1} \, \frac{\vert \vec{K} \bullet \vec{a}_\mathrm{o} \vert} {10^3 \,  \mathrm{kpc}^{-1} \, \mathrm{km}^2 \, \mathrm{s}^{-2}} + (-1)^{s_{j}} K_\mathrm{j_2} B_\mathrm{j_2}^{j_2} \frac{L}{10^8\, \mathrm{erg\,s}^{-1}} \pm 8\% 
\label{eq:or4ac} \,,
\end{eqnarray}
where $ K_\mathrm{j_2} =0.053 \pm 0.008$ and $ B_\mathrm{j_2}=2.4 \pm 0.2$ at 1$\sigma$ [for the select galaxies, \mbox{$\sigma_\mathrm{e} ({\delta}A_\mathrm{symax}/A_\mathrm{symax})=-0.04$}], for {NGC 5448} \mbox{($\delta A_\mathrm{symax} / A_\mathrm{symax} = -0.02$)}, and for {NGC 3031} \mbox{($\delta A_\mathrm{symax}/A_\mathrm{symax}  = -0.04$)}; 
\begin{eqnarray}
\frac{S_\mathrm{eor}}{10^3 \, \mathrm{kpc}^{-1} \, \mathrm{km}^2 \, \mathrm{s}^{-2}} = K_\mathrm{k_1} B_\mathrm{k_1}^{k_1} \, \frac{\vert \vec{K} \bullet \vec{a}_\mathrm{o} \vert}{10^3 \, \mathrm{kpc}^{-1} \, \mathrm{km}^2 \, \mathrm{s}^{-2}} + I_\mathrm{seor} + (-1)^{s_{k}} K_\mathrm{k_2} B_\mathrm{k_2}^{k_2} \frac{L}{10^8\, \mathrm{erg\,s}^{-1}} \pm 2\% 
\label{eq:or1ac} \,,
\end{eqnarray}
where $ K_\mathrm{k_2} =0.061 \pm 0.006$ and $ B_\mathrm{k_2}=1.80 \pm 0.05$ at 1$\sigma$; and
\begin{eqnarray}
\frac{S_\mathrm{or}}{10^3 \, \mathrm{km}^2 \, \mathrm{s}^{-2} \, \mathrm{kpc}^{-1}} = -K_\mathrm{l_1} \, B_\mathrm{l_1}^{l_1} \, \frac{L}{10^8\, \mathrm{erg\,s}^{-1}} + I_\mathrm{or} + (-1)^{s_{l}} K_\mathrm{l_2} B_\mathrm{l_2}^{l_2} \frac{\vert \vec{K} \bullet \vec{a}_\mathrm{o} \vert} {10^3 \,  \mathrm{kpc}^{-1} \, \mathrm{km}^2 \, \mathrm{s}^{-2}} \pm 2\% 
\label{eq:ksorc} \,,
\end{eqnarray}
where $ K_\mathrm{l_2} =0.0039 \pm 0.0005$ and $ B_\mathrm{l_2}=1.96 \pm 0.05$ at 1$\sigma$.  If the measured value of a parameter is larger than the first approximation calculated value, the sign $s_x =0$ in Eqs.~(\ref{eq:add1c})-(\ref{eq:ksorc}).  If the measured value of a parameter is smaller than the first approximation calculated value, the sign $s_x =1$ in Eqs.~(\ref{eq:add1c})-(\ref{eq:ksorc}).  The second approximation integer values are listed in Tables~\ref{tab:other} and \ref{tab:integer2}.

\begin{table*}
\begin{minipage}[]{140mm}
\caption{Second approximation integer values for the select galaxies. }
\label{tab:integer2} 
\centering
\begin{tabular}{lrrrrrrrrrrrrrrrrrrrrrrrr}
\hline\hline 
Galaxy&$a_2$&$b_2$&$c_2$&$d_2$&$e_2$&$f_2$&$g_2$&$h_2$&$i_2$&$j_2$&$k_2$&$l_2$\\
\hline
{NGC 0224}&&&&&&&&&&&&\\
{NGC 0598}&&&&&&&&&&&&\\
{NGC 3031}&8&1&7&6&4&4&3&6&1&4&3&11\\
{NGC 0300}&&&&&&&&&&&&\\
{NGC 2403}&3&3&4&4&6&2&4&2&5&5&5&6\\
{NGC 5457}&4&6&7&5&8&3&7&7&4&5&3&8\\
{NGC 4258}&4&2&1&3&6&2&4&3&2&4&2&1\\
{NGC 0925}&5&2&2&2&1&1&3&1&1&4&6&3\\
{NGC 2841}&&&&&&5&8&8&4&2&4&1\\
{NGC 3198}&6&6&9&1&2&5&8&6&4&3&2&7\\
{NGC 4414}&4&3&6&4&6&2&5&2&1&1&1&6\\
{NGC 3319}&4&4&6&2&4&3&7&3&5&2&6&7\\
{NGC 7331}&7&3&6&6&7&3&6&5&5&2&3&3\\
{NGC 4535}&7&6&5&4&9&3&4&2&7&4&3&6\\
{NGC 4321}&1&1&2&1&2&2&1&3&3&3&1&5\\
{NGC 4548}&16&12&8&16&21&13&14&13&16&4&5&18\\
\hline
\end{tabular}
\end{minipage}
\end{table*}

The calculated RCs of the above equations are plotted as solid lines in Fig.~\ref{fig:2} for the select galaxies. 

Because barred galaxies were part of the sample, the monopole Source suggests the $\rho$ is not sourced by matter as is gravity.  The Source appears as a monopole at the beginning of the RR that is a few kpc from the center of the galaxy.  

\section[SPM of RCs and RC asymmetry]{Discussion and conclusion}

The SPM $L$ first approximation correlation with RC parameters is an expansion of the URC concept.  The SPM includes observations such as rapidly declining RCs and asymmetric RCs.  

The second approximation equations show little effect of neighboring galaxies on the H\,{\sc i} RC except when orientation and closeness of luminous galaxies produce a higher $\vert \vec{K} \bullet \vec{a} \vert$.  Although the neighboring galaxy's effect is less significant on the RC, it does induce small perturbations in the orbits of particles that cause the observed asymmetry.  Because the $m_\mathrm{s}/m_\mathrm{\iota}$ is smaller for higher metallicity particles, the effect of neighboring galaxies on the H$\alpha$ RC is less.  Therefore, the finding of cluster effects on H$\alpha$ RCs is subject to the sample selected and may be detected only when $\vert \vec{K} \bullet \vec{a} \vert$ is large

A large $\vert \vec{K} \bullet \vec{a} \vert$ requires the host galaxy be oriented appropriately and requires large $\vec{\nabla} \rho$.  A large $\vec{\nabla} \rho $ is present near the cluster center or where the near galaxies are large and asymmetrically arranged around the host galaxy.  The former condition is caused by the $\eta < 0$ of the galaxies in the cluster center surrounded by $\epsilon >0$ of the galaxies in the cluster shell.  Therefore, the SPM is consistent with the finding of \citet{dale}.  The latter condition may be caused by a large, near neighbor galaxy.  

The SPM's $(G m_\mathrm{g} M_\mathrm{g} - G_\mathrm{s} m_\mathrm{s} L)/ m_\mathrm{\iota} R^2$ term has the form of a subtractive change in Newtonian acceleration.  The SPM suggests the $ m_\mathrm{s}$ change with radius causes the species of matter to change with radius that causes the appearance of a gravitational acceleration change with radius that is related to $L$.  If the $\vert \vec{K} \bullet \vec{a} \vert$ term is small, the SPM effective force model reduces to the MOND model as was suggested in \citet{hodg} for elliptical galaxies.

The deviation of the data of {NGC 5448} from Eq.~(\ref{eq:57ads}) suggest a physical mechanism behind the quantized galaxy parameters.  The clear departure from circular motion and the significant mass transfer inward ($\ddot{R} \neq 0$) found by \citet{fathi} suggests this galaxy is in transition from one virial state to another.  Further, the noted stellar and gas velocity difference decreases at larger radii.  The better fitting of the $v^2_\mathrm{eormax}$ -- $L$ and of the $A_\mathrm{symax}$ -- $\vert \vec{K} \bullet \vec{a}_\mathrm{o} \vert$ relations is the expected result.  {NGC 3031} shows strong, non-circular motion in the disk \citep{gott}.  This suggests the integer variation is caused by the accumulation of mass at potential barriers such as at $R_\Delta$ and $R_\mathrm{rrmax}$.  Continued nucleosynthesis and changing $\vert \vec{K} \bullet \vec{a}_\mathrm{o} \vert$ causes an occasional, catastrophic rupture of one or more of the potential barriers, $\ddot{R} \neq 0$, and, therefore, the transition of the galaxy from one integer classification to another.  A smoothly varying transition from the RR to the OR for flat or declining RCs such as {NGC 4321} suggests mass is accumulating at a potential barrier at the end of the RR and is being depleted from the outer parts of the OR.

The proposition that observed asymmetries are caused by neighboring galaxies may also apply to bars and to the formation and maintenance of the rotation of mass around the galaxy.

\citet{stei} found in a series of N-body/gas dynamical simulations that included feedback: that feedback is a necessary component for morphology determination; that the main morphological component is regulated by the mode of gas accretion and intimately linked to \emph{discrete} accretion events; that morphology is a transient phenomenon; and that the Hubble sequence reflects the varied accretion histories of galaxies.  If luminosity is proportional to the $\epsilon$, which directly causes the parameters of the RC, then there must exist a feedback mechanism controlling the parameters of the RC.

Approximately 66\% of the sample galaxies have $a_1 =4$ or $a_1=5$ as seen in Fig.~\ref{fig:3} and Table~\ref{tab:aline}.  If $a_1 =4.5$ is used in Eq.~(\ref{eq:57ads}), $v^2_\mathrm{rrmax} \propto L$ for a majority of sample galaxies.  Only {NGC 4258} of the select galaxies would appear to be an outlier, which may suggest the $v_\mathrm{rrmax}$ is larger than the measured point chosen herein.  Further, the neighboring galaxy effect would fail to improve the $v^2_\mathrm{rrmax} \propto L$ relation.  The effect of the integer values is to broaden the applicability of the parameter -- $L$ relations and to establish relations wherein the neighboring galaxy effect improves the calculation.  

The SPM does not use the mass-to-light ratio to fit RCs.  Indeed, Eqs.~(\ref{eq:57cds}) and (\ref{eq:57hds}) imply the effective mass-to-light ratio varies among galaxies as observed.

The use of the $M_\mathrm{B}$ derived from the TF and the degree of correlation achieved herein suggests the mass term is totally determined by baryonic matter.  The $\vec{\nabla} \rho$ field serves the same role in the equations as DM.  Equation~(\ref{eq:51}) with the $L$ term and Eq.~(\ref{eq:64}) with the $M_{\Delta}$ term suggests a much larger spiral galaxy baryonic mass [$M$ of Eq.~ (\ref{eq:51})] than traditional Newtonian mechanics calculates($R v^2 / G$).  Therefore, the $\vec{\nabla}\rho$ field causes Newtonian kinematic measurements to considerably underestimate mass in a galaxy \mbox{($M \approx R v^2 / G + G_\mathrm{s} m_\mathrm{s} L /G m_\mathrm{\iota}$)}.  The difference between the SPM and other RC models of the ``missing mass problem'' is that the added mass follows the luminous mass as suggested by the $ m_\mathrm{s}/m_\mathrm{\iota}$ in the equations.  Therefore, the added mass is baryonic rather than DM, a changing gravitation constant, a changing acceleration, or other form of ``fifth force''.  The ``missing mass'' is non-luminous, baryonic matter.

The purpose of the present investigation was to expand the SPM to be consistent with galaxy RC, asymmetry and EOR observations.    The resulting model adds the force of a scalar potential of the sample galaxy and of neighboring galaxies to the Newtonian rotation velocity equation.  The form of the equation for each parameter of each region is the same with differing constants.  Integer values, rather than scale factors, determine the particular galaxy parameter values.  Among the sample galaxies, the B band luminosity of a galaxy is related (1) to the maximum rotation velocity, radius, mass, and acceleration parameters at the end of the RR and at the end of the EOR and (2) to the slope of the RC in the RR and OR.  The scalar potential effect of the neighboring galaxies is related to the slope of the H\,{\sc i} RC in the EOR, to the asymmetry in the OR, and to the residual of the measured and calculated values of the above mentioned luminosity dependent parameters.  The RC in the OR may be rising, flat, or declining depending on the galaxy and its environment.  The scalar potential field causes Newtonian mechanics to considerably underestimate the mass in galaxies.  The Source of the scalar field appears as a monopole at distances of a few kpc from the center of spiral galaxies.  The resulting RC equation is consistent with the formation, evolution, and long term maintenance of asymmetry observations.  The RC observations are consistent with the SPM.

\section*{Acknowledgments}

This research has made use of the NASA/IPAC Extragalactic Database (NED) which is operated by the Jet Propulsion Laboratory, California Institute of Technology, under contract with the National Aeronautics and Space Administration.

This research has made use of the LEDA database (http://leda.univ-lyon1.fr).

I thank those people who have commented on certain aspects of this paper that needed clarification.

I acknowledge and appreciate the financial support of Maynard Clark, Apollo Beach, Florida, while I was working on this project.

\begin{appendix}

\section[SPM of RCs and RC asymmetry]{The many lines in the parameter relations are not random.}

The many lines in the plot in Fig.~\ref{fig:3} suggest the data points may be random.  The null hypothesis tested was ``the data points are indeed random points''.  This null hypothesis was tested by following the procedure used in discovering Eq.~(\ref{eq:57ads}) as follows:
(1) A trial consisted of generating 15 sets of two random numbers between zero and one and subjecting the sets to the slope, correlation coefficient, and  $\sigma_\mathrm{e}$ tests the galaxy sample passed.  Call the independent variable ($X_i$) and the dependant variable ($Y_i$), where $i$ varies from one to 15.
(2) The equation tested was
\begin{equation}
Y_{\mathrm{calc} i} = K_1 \, B^{t_i} \, X_i
\label{eq:57adt} \,,
\end{equation}
where $K_1$ is the proportionality constant, $B$ is the exponential base determined by the relation to be tested, and $t_i$ is the integer classification for the $i^\mathrm{th}$ set.
(3) The $K_1$ value for each trial was calculated as the minimum value of $K_1 = Y_i/(B\,X_i)$ with an $X_i > 0.4$.
(4) The $t_i$ value was calculated for each set of ($X_i , Y_i$) values as
\begin{equation}
t_i = \mathrm{ROUND} \left[ \log_B \, \left( \frac{Y_i}{k_1 \, X_i} \right) \right]
\label{eq:57adu} \,,
\end{equation}
where ``ROUND'' means to round the value in the braces to the nearest integer.
(5) If any subset lacked a value, that subset was ignored.
(6) If the correlation coefficient of any $t_i$ subset of ($X_i ,Y_i$) values, including the ($X_0,Y_0$) = (0,0) point, was less than 0.90, the trial failed.
(7) The $Y_{\mathrm{calc} i}$ was calculated according to Eq.~(\ref{eq:57adt}). If the $\sigma_\mathrm{e} > 0.21$ between the $Y_i$ and $Y_{\mathrm{calc} i}$ for the 15 sets, the trail failed.
(8) A count of the number $N_\mathrm{\pi}$ of sets of ($X_i , Y_i$) with $X_i^2 + Y_i^2 \leq 1$ and the number $N_\mathrm{XltY}$ of sets of ($X_i , Y_i$) with $X_i < Y_i$ was made.
(9) The trial was redone with another 15 sets of random numbers for 30,000 trials.

For $B=2 $ the results were: (1) The $N_\mathrm{\pi} /(15 \times 30,000) = 0.78528 \approx \pi/ 4$ and $N_\mathrm{XltY} /(15 \times 30,000) = 0.49999$.  Therefore, the random number generator performed satisfactorily. 
(2) Of the 30,000 trials 22 passed (0.07\%), the remainder of the trials failed.  Therefore, the null hypothesis was rejected with a 99+\% confidence.  For $B=1.65$ the confidence level of rejecting the null hypothesis was greater than 0.95.
\end{appendix}


\end{document}